# Leveraging Virtual Reality and AI Tutoring for Language Learning: A Case Study of a Virtual Campus Environment with OpenAI GPT Integration with Unity 3D

Adithya TG
Dept. Computer Science & Engineering
PES University
Bangalore, India
adithyatg.work@gmail.com

Abhinavaram N
Dept. Computer Science & Engineering
PES University
Bangalore, India
abhinava32010@gmail.com

Gowri Srinivasa
Dept. Computer Science & Engineering
PES University
Bangalore, India
gsrinivasa@pes.edu

*Abstract* — This paper presents a new approach to multiple language learning, with Hindi the language to be learn't in our case, by using the integration of virtual reality environments and AI enabled tutoring systems using OpenAI's GPT api calls. We have developed a scenario which has a virtual campus environment using Unity which focuses on a detailed representation of our university's building's 11th floor, where most of the cultural and technological activities take place. Within this virtual environment that we have created, we have an AI tutor powered by OpenAI's GPT model which was called using an api which moves around with the user. This provided language learning support in Hindi, as GPT is able to take care of language translation. Our approach mainly involves utilising speech to text, text to text conversion and text to speech capabilities to facilitate real time interaction between users and the AI tutor in the presence of internet. This research demonstrates the use of combining VR technology with AI tutoring for immersive language learning experiences and provides interaction.

*Keywords—Virtual Reality, AI Tutoring, GPT, Language Learning, Speech Recognition, Text to Speech*

## I. INTRODUCTION

Language learning is a complex process that usually requires a lot of studies and personalised guidance to learn language. The usual normal, traditional methods often lacks continuous interaction and fails to provide the one who is learning a particular language with authentic language practice opportunities. Using virtual reality is now a promising solution by which, immersive environments are created in which learners can interact with virtual scenarios of their available choice and engage in real world language interactions. Also, AI powered tutoring systems have shown a high potential in providing personalised language learning assistance at the own pace of each and every user.

In this paper, we present a case study of integrating virtual reality technology with an AI tutoring system for language learning. We created a detailed virtual campus environment and integrated an AI tutor powered by OpenAI's GPT model. We focused on providing language learning support in Hindi, leveraging speech to text, text to text and text to speech features to give real time interaction between users and the AI tutor.

## II. RELATED WORKS

There are multiple works that give a detailed review on the usage of Virtual Reality in language learning. These studies identify high-impact journals in the field of Computer-Assisted Language Learning (CALL) and systematically screen studies published between 2010 and 2020[1][2]. Additionally, there is research examining 69 empirical research articles selected for their relevance to VR-supported language learning. This research involves two stages: preliminary coding and analytical stages, followed by formal coding and analytical stages, using a coding scheme and analytical framework proposed by previous studies to address research questions related to VR-supported language learning [2][3]. Another research highlights gaps in the existing scientific literature on virtual reality in language learning, emphasizing the need for continued research and development in this field to improve language acquisition techniques and methods. The study also provides insights into the potential benefits and challenges of incorporating VR into language teaching and learning practices [4].

Moreover, an article by the Royal Society explores how GPT can analyze text data on augmented reality interfaces. It notes that GPT-4, a paid and improved model, provides better summaries than the commonly available free GPT-3.5 Turbo. Participants testing the models had mixed feelings about the interfaces, with some finding them clear and others confusing. GPT-4 highlighted issues and offered improvement suggestions, concluding that GPT was more useful for analyzing user feedback in human-computer interaction research than existing traditional methods. This led to the use of GPT's API for language translations [5].

When considering the text-to-speech (TTS) capabilities of GPT and Bard for use, each platform has its own strengths and limitations. GPT, created by OpenAI, currently lacks a native TTS feature, requiring users to use third-party tools to convert text to speech. This offers flexibility in selecting TTS voices but adds an extra step and potential extra costs. On the other hand, Bard, Google's AI chatbot, has integrated TTS that can promptly transform generated text into speech, making it convenient for users to listen to the content. Bard's TTS is noted for its quality speech synthesis, enhancing the user experience. Therefore, choosing between GPT or Bard for TTS in research paper writing depends on user preferences, with GPT offering more control over TTS through third-party integrations and Bard providing a more integrated and user-friendly TTS solution.

Additionally, a paper discusses the development of a smart multiplatform game application using the UNITY3D engine aimed at enhancing CPR education. Recognizing the critical importance of CPR skills in emergency situations, the study emphasizes the need for effective training methods that engage learners. Traditional CPR training often lacks interactivity and can be perceived as tedious, which may hinder retention of skills. To address this, the authors propose a gamified approach that incorporates elements of play and competition, making the learning process more enjoyable and memorable. The application is designed to be accessible across various platforms, ensuring a wider reach and facilitating learning anytime and anywhere. The integration of multimedia elements and interactive scenarios aims to simulate real-life situations, allowing users to practice and refine their CPR skills in a safe environment. The findings suggest that such educational tools could play a significant role in public health initiatives by improving knowledge retention and increasing the likelihood of individuals performing CPR in emergencies [15].

Furthermore, research titled "Research on The Application of AR Technology Based on Unity3D in Education" by ChangYuan Li and BaiHui Tang investigates the integration of Augmented Reality (AR) technology in educational contexts, emphasizing its immersive and interactive capabilities. The authors present a comprehensive display system designed to demonstrate human body tissues and structures, developed using Unity3D and Vuforia. This system is packaged as an Android application, showcasing strong user engagement and interaction. The study highlights the effectiveness of a multi-interactive and cross-platform AR technology scheme, achieving an impressive recognition accuracy rate of over 98% for the application. The system comprises three main modules: interaction, user interface (UI), and sound, facilitating seamless user interaction with the educational content. The findings suggest that the Vuforia+Unity3D AR technology is highly adaptable, allowing for easy content replacement to create various educational demonstrations. This research underscores the significant potential of AR technology in enhancing educational experiences, providing a strong foundation for future applications in teaching and learning environments [14].

## III. METHODOLOGY

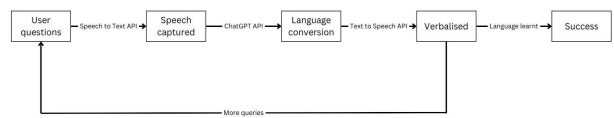

Fig 1. Workflow of the model

Fig 1 represents the block diagram simulation of overall working of the model.

A. Virtual Campus Environment Development

Scenario Design: An environment for a particular scenario that the user wants to be trained or wants an experience will be chosen by the user virtually which might be a market scenario, classroom scenario, festive scenario and what not. In our case, we developed a culturally vibrant floor of our college as seen in Fig 2, where Bharatanatyam is being performed by a few in built models as seen in Fig 3 and Fig 5, there's a musical instrument's room as seen in Fig 4 and Fig 6 and multiple other rooms. Users are given the option of using a VR headset to move around and interact or use the keyboard and mouse on the Unity3D app to interact with various objects in the space, use and interact with them as and when needed by them.

Traditional development: In the paper, Unity3D is used, which is a popular game development platform, to create a virtual college platform of the 11th floor. We worked with Blender also, to create assets and information in the scene so it looked realistic and added animations to it. Basically, focused on creating the 11th floor of the university, which included a detailed architecture and multiple interactive features which can be accessed by the user who is interacting with the scene, who wants to learn about anything in the scene.

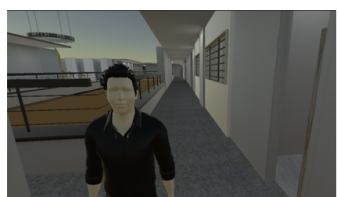 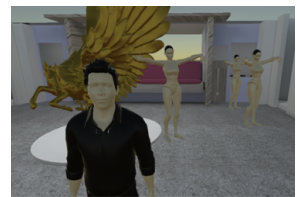

Fig 2. Initial screen    Fig 3. User Near Bharatanatyam dancers

B. AI Tutor Integration

OpenAI GPT: OpenAI's GPT3.5 Turbo model was used in the virtual environment to provide language learning assistance, which converted the received text from the user into the language for their choice, in our case was Hindi through API calls. GPT is considered to be a state of the art language model, which is very well capable of generating human like text responses and is quite good at it.

API Integration: We made API calls to OpenAI's API service to interact with the GPT3.5 Turbo model, which had a limitation of high speed internet but was taken care of. Speech inputs from users are converted to text and sent as prompts to the model along with a few more additional prompts, by which it generates text responses in Hindi.

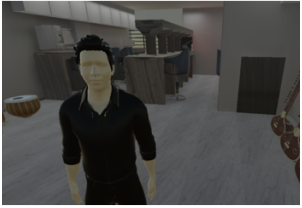

Fig 4. User near musical instruments

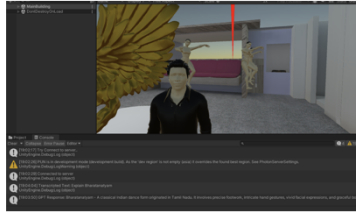

Fig 5. Explanation when near Bharatanatyam

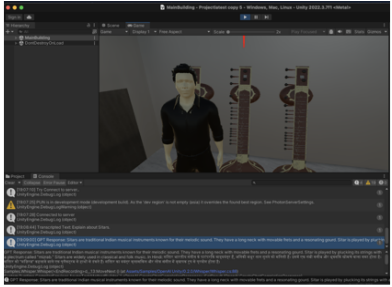

Fig 6. Audio captured from speech and text generated near Bharatanatyam dancer along with prompt

## C. Speech to text and text to speech

Speech Recognition: To begin with, the Whisper model is used here. It is a text to speech model that employs a sophisticated and novel approach to converting text into speech by combining several key components of the speech production process into a unified architecture. High quality speech recognition and transcription are accomplished using the Whisper text to speech (TTS) model by OpenAI that leverages a promising encoder decoder transformer setting. The model is trained with publicly available multilingual audio and text data on the web with a total of 680 thousand hours. The Whisper model begins with segmenting the input audio into individual 30 second chunks. It transforms each of these audio segments into a mel frequency cepstrum (MFC) – a feature that is quite resistant to noise and accents in the sound. These features of the MFC are then fed into the encoder section of the transformer model. The encoder part of the model takes the audio features of the input and encodes them into a sequence of representations. The decoder component in the transformer model is then learned to predict the transcription of each audios segment using sequence to sequence learning. The decoder can also be instructed to perform other functions such as recognising the language being used or translating speech interpretation. This is because Whisper has had access to a massive training corpus and can be fine tuned to adapt to various tasks with high accuracy, especially in English speech recognition. The whole span architecture of the model eliminates the use of several specific components that are used in a standard pipeline for speech processing. This makes Whisper easy to use and allows developers to quickly plug in access to very accurate and efficient speech to text functionality anywhere. Therefore, OpenAI's speech to text API which uses the Whisper model is used so that the words spoken by the user are read and converted to text. The API call is done when the user clicks on an object or goes close to it. There is an option to speak to the model directly to ask questions in general also, even if not near to any object. This is limited to internet connectivity for faster and realtime responses.

Text to Speech Conversion: Text responses generated and converted by OpenAI's GPT from English to Hindi is to be read aloud, which is done by OpenAI's text to speech API. Using this, the user is made possible to hear the AI tutor's responses in natural sounding Hindi speech, which helps users to understand and learn a new language as shown in Fig 5 and 6, with a detailed overview of the conversations recorded in Fig 7 and 8 and table 2. This again, is limited as to internet connectivity for faster and realtime responses.

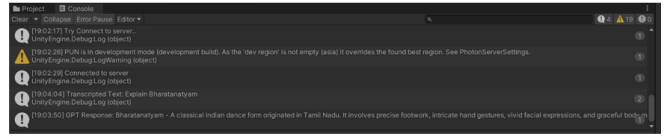

Fig 7. Conversation with AI chatbot, providing text and speech in Hindi as well as English

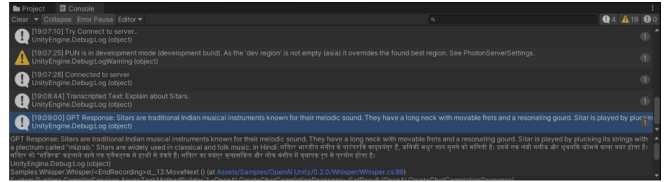

Fig 8. Audio captured from speech and text generated near Sitar along with user query

## IV. RESULTS

Integration of Virtual Reality technology with AI tutoring yielded promising results in providing a good different language learning experience. Real time interaction with the AI tutor allowed users to practise language skills in a dynamic and interactive manner which was made possible by the help of good internet connectivity.

Further more, the model, on a dataset of the theme of education and motivation, obtained an average accuracy of 90.59% in accurate text generations.

## V. OBJECTIVE EVALUATION

The model was also evaluated using the BLEU scores for machine language translations of 100 different questions. The BLEU score in simple terms, is the standard that machine translation quality is measured on. It is used to compare a reference human translation to machine translation on a scale of 0 to 100.

The model for translating English to Hindi, obtained a good BLEU score of 71.08, indicating resemblance between the actual meaning and the translated text meaning.

## VI. OBJECTIVE EVALUATION

The model was tested by 20 different users, with 18 of them reported that it had a high level of engagement and were satisfied with the virtual campus environment and the AI tutor's support to learn a new language, which was Hindi in our case. They rated the model and the environment on the basis of Accuracy of Translation, Ease of Use, Comprehensiveness, Engagement, Learning Outcome and Overall Satisfaction on a scale of 1 to 5 as shown in table 1.

Some of the questions asked by users and the responses generated by the AI model are mentioned in table 2 along with their translations.

After the user's query is received by the model, it generates an English response and a Hindi response. The

| User No | Accuracy of Translation | Ease of Use | Comprehensiveness | Engagement | Learning Outcome | Overall Satisfaction |
|---|---|---|---|---|---|---|
| 1 | 4 | 4 | 5 | 3 | 4 | 4 |
| 2 | 3 | 5 | 5 | 4 | 4 | 4 |
| 3 | 4 | 5 | 5 | 4 | 5 | 5 |
| 4 | 4 | 3 | 2 | 3 | 3 | 3 |
| 5 | 3 | 4 | 4 | 3 | 4 | 3 |
| 6 | 4 | 3 | 5 | 4 | 3 | 4 |
| 7 | 2 | 2 | 4 | 3 | 2 | 2 |
| 8 | 5 | 3 | 3 | 3 | 4 | 4 |
| 9 | 3 | 3 | 3 | 4 | 4 | 3 |
| 10 | 4 | 4 | 4 | 3 | 3 | 4 |
| 11 | 3 | 5 | 3 | 3 | 4 | 4 |
| 12 | 4 | 4 | 3 | 3 | 5 | 4 |
| 13 | 3 | 3 | 4 | 3 | 4 | 3 |
| 14 | 3 | 4 | 2 | 3 | 2 | 2 |
| 15 | 3 | 4 | 4 | 5 | 3 | 4 |
| 16 | 4 | 5 | 4 | 4 | 4 | 5 |
| 17 | 3 | 4 | 3 | 5 | 5 | 5 |
| 18 | 4 | 3 | 3 | 4 | 3 | 3 |
| 19 | 4 | 4 | 4 | 5 | 4 | 5 |
| 20 | 5 | 4 | 4 | 5 | 5 | 5 |

Table 1. Reviews of multiple users

English query is first readout, followed by the Hindi response, where the previous query is also stored and related questions can be asked back to the model, reducing hallucinations and repetition of content.

| Quer No | English Query | English Response | Hindi Response |
|---|---|---|---|
| 1 | Tell me something about the Indian instrument Sitar | The sitar is a classical Indian string instrument, known for its distinctive sound and use in Hindustani classical music. It has a long neck, a gourd resonator, and multiple strings. | सितार एक शास्त्रीय भारतीय तार वाद्य यंत्र है, जो अपनी विशिष्ट ध्वनि और हिंदुस्तानी शास्त्रीय संगीत में उपयोग के लिए जाना जाता है। इसमें लंबी गर्दन, लौकी का गुंजक और कई तार होते हैं। |
| 2 | Tell me something about the Indian dance form Bharatanatyam | Bharatanatyam is a classical Indian dance form originating from Tamil Nadu. It combines expressive hand gestures, footwork, and facial expressions to convey stories, often based on Hindu mythology. | भरतनाट्यम तमिलनाडु से उत्पन्न एक शास्त्रीय भारतीय नृत्य है। यह कहानियों बताने के लिए हाथ के हाव-भाव, पैर की चाल और चेहरे के भावों को मिलाता है, जो अक्सर हिंदू पौराणिक कथाओं पर आधारित होती है। |
| 3 | Give me tips to be a public speaker | Practice regularly, know your audience, structure your speech, use visuals, engage with stories, maintain eye contact, manage nervousness, use body language, seek feedback, stay authentic. | नियमित अभ्यास, श्रोताओं को जानें, भाषण का संरचना, दृश्य सामग्री, कहानियों से जुड़ें, आँखों से संपर्क, नर्वसनेस का नियंत्रण, शारीरिक भाषा, प्रतिक्रिया प्राप्त करें, प्रामाणिक रहें। |
| 4 | What could I speak about working hard | Working hard is essential for achieving success. It requires dedication, persistence, and a strong work ethic. When you put in the effort consistently, you increase your chances of reaching your goals. | कड़ी मेहनत करना सफलता पाने के लिए अत्यंत आवश्यक है। इसमें समर्पण, दृढ़ इच्छाशक्ति और मजबूत कामयाबी का अभ्यास होता है। जब आप निरंतर प्रयास करते हैं, तो आप अपने लक्ष्यों तक पहुँचने की संभावनाएँ बढ़ाते हैं। |

Table 2. English Queries with their responses and their translated responses.

We can arrive at a decent conclusion that the idea of using Augmented reality for learning is quite helpful as seen from table 1. Since the GPT3.5-turbo is good enough in converting English responses to Hindi responses, it would be an ideal, state of the art model to use in the Unity3D environment.

## VII. DISCUSSIONS AND FUTURE WORKS

The idea and implementation shows the integration of a Virtual Reality environment, college 11th floor with an AI tutoring system, basically a voiced chatbot for language learning. Future work will mainly focus on expanding language support by including multiple additional languages and adding different kinds of dialects. Addition of multiplayer support and gamified learning experiences by creating a points based system, is also something to be considered. Also the future work is on the creation of multiple scenarios and the synthetic creation of these scenarios as and when required by the user for their choice of liking.

## VIII. CONCLUSION

In conclusion, our work shows the novel idea of combining virtual reality with AI tutoring with the help of a voiced chatbot for language learning. The integration of a virtual campus environment with an AI tutor created with the help of OpenAI APIs creates an approach to providing a captivating and individualised, own paced language learning experience for each and every particular user. This research mainly contributes to the advancement of a new solution for language education and has detailed design for future virtual reality scenario based learning environments.